\documentclass[twocolumn]{aastex631}
\usepackage{amsmath}
\usepackage{color}
\usepackage{url}
\usepackage{ulem}
\usepackage{multirow}
\usepackage{amsmath}
\usepackage{wasysym}

\usepackage{rotating, graphicx}

\def\apjl{ApJ}%
\def\apjs{ApJS}%
\def\ao{Appl.~Opt.}%
\def\aap{A\&A}%
\newcommand\C   {\textsc{Clumpy}}

\shorttitle{BASS XLII: the covering factor of dusty gas in nearby AGN}
\shortauthors{Ricci et al.}

\begin{document}

\title{BASS XLII: The relation between the covering factor of dusty gas and the Eddington ratio in nearby active galactic nuclei}

\author[0000-0001-5231-2645]{C. Ricci}
\affiliation{Instituto de Estudios Astrof\'{\i}sicos, Facultad de Ingenier\'{\i}a y Ciencias, Universidad Diego Portales, Avenida Ejercito Libertador 441, Santiago, Chile}
\affiliation{Kavli Institute for Astronomy and Astrophysics, Peking University, Beijing 100871, China}
\author[0000-0002-4377-903X]{K. Ichikawa}
\affiliation{Frontier Research Institute for Interdisciplinary Sciences, Tohoku University, Sendai, 980-8578, Japan}
\author[0000-0001-5146-8330]{M. Stalevski}
\affiliation{Astronomical Observatory, Volgina 7, 11060 Belgrade, Serbia}
\affiliation{Sterrenkundig Observatorium, Universiteit Ghent, Krijgslaan 281 S9, B-9000 Ghent, Belgium}
\author[0000-0002-6808-2052]{T. Kawamuro}
\affiliation{RIKEN Cluster for Pioneering Research, 2-1 Hirosawa, Wako, Saitama 351-0198, Japan}
\author[0000-0002-9754-3081]{S. Yamada}
\affiliation{RIKEN Cluster for Pioneering Research, 2-1 Hirosawa, Wako, Saitama 351-0198, Japan}
\author[0000-0001-7821-6715]{Y. Ueda}
\affiliation{Department of Astronomy, Kyoto University, Kitashirakawa -Oiwake-cho, Sakyo-ku, Kyoto 606-8502, Japan}
\author[0000-0002-7962-5446]{R. Mushotzky}
\affiliation{Department of Astronomy, University of Maryland, College Park, MD 20742, USA}
\affiliation{Joint Space-Science Institute, University of Maryland, College Park, MD 20742, USA}
\author[0000-0003-3474-1125]{G. C. Privon}
\affiliation{National Radio Astronomy Observatory, 520 Edgemont Road, Charlottesville, VA 22903, USA}
\affiliation{Department of Astronomy, University of Florida, P.O. Box 112055, Gainesville, FL 32611, USA}
\author[0000-0002-7998-9581]{M. J. Koss}
\affiliation{Eureka Scientific, 2452 Delmer Street Suite 100, Oakland, CA 94602-3017, USA}
\affiliation{Space Science Institute, 4750 Walnut Street, Suite 205, Boulder, CO 80301, USA}
\author[0000-0002-3683-7297]{B. Trakhtenbrot}
\affiliation{School of Physics and Astronomy, Tel Aviv University, Tel Aviv 69978, Israel}
\author[0000-0002-9378-4072]{A. C. Fabian}
\affiliation{Institute of Astronomy, Madingley Road, Cambridge CB3 0HA, UK}
\author[0000-0001-6947-5846]{L. C. Ho}
\affiliation{Kavli Institute for Astronomy and Astrophysics, Peking University, Beijing 100871, China}
\affiliation{Department of Astronomy, School of Physics, Peking University, Beijing 100871, China}
\author{D. Asmus}
\affiliation{Department of Physics \& Astronomy, University of Southampton, Southampton, SO17 1BJ, UK}
\affiliation{Gymnasium Schwarzenbek, 21493 Schwarzenbek, Germany}
\author[0000-0002-8686-8737]{F. E. Bauer}
\affiliation{Instituto de Astrof\'isica, Facultad de F\'isica, Pontificia Universidad Cat\'olica de Chile, Campus San Joaquin, Av. Vicu\~na Mackenna 4860, Macul Santiago, Chile, 7820436}
\affiliation{Centro de Astroingenier\'ia, Facultad de F\'isica, Pontificia Universidad Cat\'olica de Chile, Campus San Joaquin, Av. Vicu\~na Mackenna 4860, Macul Santiago, Chile, 7820436}
\affiliation{Millennium Institute of Astrophysics, Nuncio Monse\~nor S\'otero Sanz 100, Of 104, Providencia, Santiago, Chile}
\author[0000-0001-9910-3234]{C. S. Chang}
\affiliation{Joint ALMA Observatory, Avenida Alonso de Cordova 3107, Vitacura 7630355, Santiago, Chile}
\author[0009-0007-9018-1077]{K. K. Gupta}
\affiliation{STAR Institute, Quartier Agora - All\'ee du six Ao\^ut, 19c B-4000 Li\`ege, Belgium}
\author[0000-0002-5037-951X]{K. Oh}
\affiliation{Korea Astronomy \& Space Science institute, 776, Daedeokdae-ro, Yuseong-gu, Daejeon 34055, Republic of Korea}
\author[0000-0003-2284-8603]{M. Powell}
\affiliation{Kavli Institute for Particle Astrophysics and Cosmology, Stanford University, 452 Lomita Mall, Stanford, CA 94305, USA}
\affiliation{Department of Physics, Stanford University, 382 Via Pueblo Mall, Stanford, CA 94305, USA}
\author[0000-0001-8640-8522]{R. W. Pfeifle}
\affiliation{X-ray Astrophysics Laboratory, NASA Goddard Space Flight Center, Code 662, Greenbelt, MD 20771, USA}
\affiliation{Oak Ridge Associated Universities, NASA NPP Program, Oak Ridge, TN 37831, USA}
\author[0000-0003-0006-8681]{A. Rojas}
\affiliation{Centro de Astronom\'{\i}a (CITEVA), Universidad de Antofagasta, Avenida Angamos 601, Antofagasta, Chile}
\affiliation{Instituto de Estudios Astrof\'{\i}sicos, Facultad de Ingenier\'{\i}a y Ciencias, Universidad Diego Portales, Avenida Ejercito Libertador 441, Santiago, Chile}
\author[0000-0001-5742-5980]{F. Ricci}
\affiliation{Dipartimento di Matematica e Fisica, Universita Roma Tre, via della Vasca Navale 84, I-00146, Roma, Italy}
\author[0000-0001-8433-550X]{M. J. Temple}
\affiliation{Instituto de Estudios Astrof\'{\i}sicos, Facultad de Ingenier\'{\i}a y Ciencias, Universidad Diego Portales, Avenida Ejercito Libertador 441, Santiago, Chile}
\author[0000-0002-3531-7863]{Y. Toba}
\affiliation{National Astronomical Observatory of Japan, 2-21-1 Osawa, Mitaka, Tokyo 181-8588, Japan}
\affiliation{Academia Sinica Institute of Astronomy and Astrophysics, 11F Astronomy-Mathematics Building, AS/NTU, No.1, Section 4, Roosevelt Road, Taipei 10617, Taiwan}
\affiliation{Research Center for Space and Cosmic Evolution, Ehime University, 2-5 Bunkyo-cho, Matsuyama, Ehime 790-8577, Japan}
\author[0000-0003-3450-6483]{A. Tortosa}
\affiliation{INAF - Osservatorio Astronomico di Roma, via di Frascati 33, 00078 Monte Porzio Catone, Italy}
\author[0000-0001-7568-6412]{E. Treister}
\affiliation{Instituto de Astrof\'isica, Facultad de F\'isica, Pontificia Universidad Cat\'olica de Chile, Campus San Joaquin, Av. Vicu\~na Mackenna 4860, Macul Santiago, Chile, 7820436}
\author{F. Harrison}
\affiliation{Cahill Center for Astronomy and Astrophysics, California Institute of Technology, Pasadena, CA 91125, USA}
\author[0000-0003-2686-9241]{D. Stern}
\affiliation{Jet Propulsion Laboratory, California Institute of Technology, 4800 Oak Grove Drive, MS 169-224, Pasadena, CA 91109, USA}
\author[0000-0002-0745-9792]{C. M. Urry}
\affiliation{Yale Center for Astronomy \& Astrophysics, Physics Department, PO Box 208120, New Haven, CT 06520-8120, USA}

\correspondingauthor{Claudio Ricci}
\email{claudio.ricci@mail.udp.cl}

\begin{abstract}
Accreting supermassive black holes (SMBHs) located at the center of galaxies are typically surrounded by large quantities of gas and dust. The structure and evolution of this circumnuclear material can be studied at different wavelengths, from the submillimeter to the X-rays. Recent X-ray studies have shown that the covering factor of the obscuring material tends to decrease with increasing Eddington ratio, likely due to radiative feedback on dusty gas. Here we study a sample of 549 nearby ($z\lesssim 0.1$) hard X-ray (14--195\,keV) selected non-blazar active galactic nuclei (AGN), and use the ratio between the AGN infrared and bolometric luminosity as a proxy of the covering factor. We find that, in agreement with what has been found by X-ray studies of the same sample, the covering factor decreases with increasing Eddington ratio. We also confirm previous findings which showed that obscured AGN typically have larger covering factors than unobscured sources. Finally, we find that the median covering factors of AGN located in different regions of the column density-Eddington ratio diagram are in good agreement with what would be expected from a radiation-regulated growth of SMBHs.
\end{abstract}	
               

\setcounter{footnote}{0}

\section{Introduction}

Supermassive black holes (SMBHs) that accrete copious amounts of gas and dust from their surroundings can emit throughout the whole electromagnetic spectrum, and are observed as Active Galactic Nuclei (AGN). The material responsible for most of the AGN obscuration is widely believed to be anisotropically distributed, possibly in the form of a torus, as predicted by the classical AGN unification model \citep{Antonucci:1993fu,Urry:1995le,Netzer:2015rev,Ramos-Almeida:2017fv,Hickox:2018qy}. This obscuring material can be studied in the X-ray band in absorption (e.g., \citealp{Awaki:1991kq,Ueda:2003qf,Merloni:2014qv,Ricci:2015tg}) and/or through spectral features (e.g., \citealp{Lightman:1988sp,Pounds:1990hb,Matt:1991gd}), in the infrared (IR) from the radiation produced by reprocessing of the UV/optical and X-ray radiation in circumnuclear dust (e.g., \citealp{Krolik:1988kk,Granato:1994pb,Jaffe:2004la,Elitzur:2008ec,Gandhi:2009uq,Ramos-Almeida:2011eb,Alonso-Herrero:2011jx,Lanz:2019zd,Gamez-Rosas:2022gp}), or at millimeter wavelengths from dust continuum and molecular line emission (e.g., \citealp{Impellizzeri:2019lm,Garcia-Burillo:2019qx,Garcia-Burillo:2021qc,Imanishi:2020er,Tristram:2022st}).  Understanding the structure and evolution of the gas and dust surrounding these accreting SMBHs is extremely important, because this is likely to be the reservoir of material that eventually accretes onto SMBHs. Moreover, being located in the inner regions of the accreting system, this material could carry imprinted signatures of AGN feedback (e.g., \citealp{Fabian:2006lq,Ricci:2017ss}), which is thought to play an important role in the evolution of galaxies (e.g., \citealp{Kormendy:2013af,Harrison:2017nj} and references therein). 

One of the main parameters of the obscuring material surrounding the SMBH is its covering factor (e.g., \citealp{Lawrence:2010ov}), i.e. the fraction of the SMBH sky that is obscured, which can be inferred using the following techniques in the X-rays, optical and IR: \newline
i) In X-ray surveys using the {\it fraction of obscured sources} ($f_{\rm obs}$; e.g., \citealp{Ueda:2003qf,Ueda:2014ix,La-Franca:2005uf,Hasinger:2008ve,Merloni:2014qv}), i.e. the fraction of AGN with column densities $\log (N_{\rm H}/\rm cm^{-2})\geq 22$. According to the unification model, different objects would probe different inclination angles with respect to the torus. Therefore, if the sample has a high completeness level, then the fraction of sources within a certain range of column densities is a proxy of the mean covering factor of the obscuring material with that column density. Alternatively, the fraction of AGN optically classified as type-1 (i.e., showing both broad permitted and narrow forbidden optical lines) and type-2 (displaying only narrow lines) can also be used to infer the covering factor of the obscuring material (e.g., \citealp{Lawrence:1982ys,Simpson:2005uu,Toba:2013jf,Oh:2015if}). However, a possible source of uncertainty related to this approach is the existence of a fraction of AGN which are optically dull (e.g., \citealp{Smith:2014pf,Georgantopoulos:2005cp,Koss:2017fp}), and of unobscured sources in which the optical broad lines are too faint or broad to be detected (e.g., \citealp{Bianchi:2017oj}).\newline
ii) From the {\it ratio between the IR and the bolometric AGN luminosity}. Considering that a significant fraction of the IR emission in AGN is produced by dust reprocessing, it has been argued that the fraction of bolometric luminosity ($L_{\rm Bol}$) re-emitted in the IR is directly proportional to the covering factor of the obscuring material (e.g., \citealp{Maiolino:2007ii,Treister:2008ff,Gandhi:2009uq,Assef:2013bu,Netzer:2016aa,Toba:2021ys}). However, a detailed study carried out using radiative transfer simulations of dusty tori \citep{Stalevski:2016kl} has shown that the relation between the covering factor and the ratio between the IR and bolometric AGN luminosity is significantly more complex than previously believed (see \S\ref{sect:CFdust}).\newline
iii) From torus models applied to the {\it mid-IR spectral energy distribution} (SED) of individual AGN. Several IR torus models have been developed in the past decade \citep{Honig:2006tg,Schartmann:2008kv,Honig:2010tt,Honig:2010vp,Honig:2017qp,Stalevski:2012kq,Stalevski:2016kl,Siebenmorgen:2015br}, and can be used to recover some of the properties of the dust surrounding SMBHs, including its covering factor. Works carried out using torus models such as \textsc{clumpy} \citep{Nenkova:2008lg,Nenkova:2008kl,Nikutta:2009qy}, \textsc{cat3d} \citep{Honig:2010tt} and \textsc{cat3d-wind} \citep{Honig:2017qp} have been applied to a significant number of AGN (e.g., \citealp{Mor:2009fk,Alonso-Herrero:2011jx,Alonso-Herrero:2021id,Ramos-Almeida:2011eb,Ichikawa:2015qq,Zhuang:2018eg,Gonzalez-Martin:2019gy,Garcia-Bernete:2019mp,Garcia-Bernete:2022pk}). For one of the closest AGN, the Circinus Galaxy, it has been possible to infer the covering factor by modelling both spectroscopic and high-resolution morphological data \citep{Stalevski:2017qv,Stalevski:2019gl}.\newline
iv) Applying torus models to the {\it broad-band X-ray spectra of AGN}. X-ray torus models (e.g., \citealp{Murphy:2009ly,Brightman:2011fe,Liu:2014qy,Paltani:2017zt,Balokovic:2018lk,Tanimoto:2019ts,Buchner:2021kv,Ricci:2023lp}), some of which now include the effect of dusty gas in the X-rays \citep{Ricci:2023lp,Vander-Meulen:2023uq}, allow us to infer the covering factor of the gas and dust surrounding the SMBH. This has been particularly effective in the past years using {\it NuSTAR} observations (e.g., \citealp{Brightman:2015cr,Zhao:2020ql,Tanimoto:2020lh,Ogawa:2019jz,Ogawa:2021um,Yamada:2021il,Uematsu:2021nr,Inaba:2022uo,Andonie:2022qx}), since its large energy coverage ($3-79$\,keV) allows for the detection of both main features produced by reprocessed X-ray radiation: the Fe\,K$\alpha$ line (6.4\,keV) and the Compton hump ($\sim 30$\,keV).

A relation between the torus covering factor and the AGN luminosity was originally found by the decrease in the fraction of type-2 AGN with the luminosity \citep{Lawrence:1982ys,Lawrence:1991vn}. Studies of X-ray (e.g., \citealp{Ueda:2003qf,Ueda:2014ix,Treister:2005qi,Treister:2006hj,Sazonov:2007if,Della-Ceca:2008xu,Hasinger:2008ve,Beckmann:2009fk,Ueda:2011fk,Merloni:2014qv,Buchner:2015ve}) and IR (e.g., \citealp{Maiolino:2007ii,Treister:2008ff,Gandhi:2009uq,Assef:2013bu,Lusso:2013fy,Toba:2014ao,Lacy:2015rm,Stalevski:2016kl,Mateos:2016ys,Ichikawa:2017xr}) surveys, as well as IR studies carried out by applying torus models (e.g., \citealp{Mor:2009fk,Alonso-Herrero:2011jx}) have confirmed this trend. The relationship between the covering factor of the obscuring material and the luminosity has been shown to be related to the intrinsically different luminosity functions of obscured and unobscured AGN (e.g., \citealp{Tueller:2008yg,Della-Ceca:2008xu,Burlon:2011dk}), and that it can reproduce the decrease of the Fe\,K$\alpha$ intensity with luminosity (\citealp{Ricci:2013cz}, see also \citealp{Iwasawa:1993yb,Bianchi:2007et,Ricci:2014dg,Boorman:2018fj,Matt:2019nq}). Recently, it has been shown that the main driver of this correlation could be the Eddington ratio (\citealp{Ricci:2017ss,Ricci:2022xp}, see also \citealp{Ezhikode:2017tu,Buchner:2017gf,Ananna:2022tr}), with radiation pressure on dusty gas (e.g., \citealp{Fabian:2006lq,Venanzi:2020dd}) likely responsible for the decrease of the covering factor with increasing mass-normalized accretion rates. Here we complement our previous X-ray studies \citep{Ricci:2017ss,Ricci:2022xp} by analyzing the relation between the dust covering factor, obtained by comparing IR to bolometric luminosities, and the Eddington ratio for sources from the BAT AGN Spectroscopic survey (BASS\footnote{www.bass-survey.com}, \citealp{Koss:2017fp,Koss:2022yy,Ricci:2017cj}). Throughout the paper we adopt standard cosmological parameters ($H_{0}=70\rm\,km\,s^{-1}\,Mpc^{-1}$, $\Omega_{\mathrm{m}}=0.3$, $\Omega_{\Lambda}=0.7$).

\section{Sample and Data}\label{sect:sampledata}

Our sample is composed of objects from the BASS survey. BASS is studying in detail the multi-wavelength properties of AGN detected by the Burst Alert Telescope (BAT, \citealp{Barthelmy:2005uq}), on board the {\it Neil Gehrels Swift Observatory} \citep{Gehrels:2004kx}, with the goal of improving our understanding of the properties of accreting SMBHs in the local Universe ($z\lesssim 0.1$). {\it Swift}/BAT operates in a band (14--195\,keV) that is not significantly affected by obscuration up to column densities $\log (N_{\rm H}/\rm cm^{-2})\simeq 24$ (e.g., Fig.\,1 in \citealp{Ricci:2015tg}, see also \citealp{Koss:2016kq}), thus providing an almost unbiased view of local AGN. {\it Swift}/BAT has detected 733 non-blazar AGN in the first 70-months of operations \citep{Baumgartner:2013ee,Ricci:2017cj}, and our initial sample consists of the 731 sources for which X-ray spectroscopy was available. The intrinsic 14--150\,keV AGN luminosities ($L_{14-150}$) and their column densities are taken from \cite{Ricci:2017cj}, which reports the broadband (0.3--150\,keV) X-ray spectral properties of the AGN from the {\it Swift}/BAT 70-month catalog. The 1-1000\,$\mu$m luminosity of the torus ($L_{\rm Tor}$), necessary to estimate its covering factor (\S\ref{sect:CFdust}), was calculated from the 12\,$\mu$m AGN luminosity ($L_{12\mu\rm m}$) reported in \citet{Ichikawa:2019zz}, using a correction factor of $\kappa_{\rm IR}=2.92$ (i.e., $L_{\rm Tor}=\kappa_{\rm IR}\times L_{12\mu\rm m}$), based on the model of \cite{Stalevski:2016kl}. The AGN IR luminosities of \citet{Ichikawa:2019zz} were obtained by the spectral decomposition of the IR SED, considering both AGN and host galaxy emission using \textsc{decompir} \citep{Mullaney:2011ei}, and are shown to be in good agreement with the luminosities obtained by high-spatial-resolution IR studies (e.g., \citealp{Asmus:2014oq}, see Fig.\,11 of \citealp{Ichikawa:2019zz}). \citet{Ichikawa:2019zz} report IR luminosities for 587 non-blazar AGN, of which 21 only have upper limits, since their IR SED is dominated by other physical processes, such as star formation.

The black hole masses ($M_{\rm BH}$) for the sources of the BASS sample are listed in \cite{Koss:2017fp} and \cite{Koss:2022bb}, who reported black hole masses for 790 AGN from the {\it Swift}/BAT 70-month catalog \citep{Baumgartner:2013ee}. For unobscured AGN black hole masses were mostly estimated using broad H$\alpha$ and H$\beta$ lines, while for obscured objects we used black hole masses mostly estimated with velocity dispersion (see \citealp{Koss:2022bb,Koss:2022vc} for details). Similarly to what was done in \citet{Ricci:2017ss}, we excluded the objects with $N_{\rm H}\geq 10^{22}\rm\,cm^{-2}$ for which $M_{\rm BH}$ was obtained using broad H$\alpha$ and H$\beta$, since the black hole masses are likely underestimated because of the extinction of the optical emission (e.g., \citealp{Ricci:2022sg,Mejia-Restrepo:2022sd}). Cross-matching the 587 AGN with $L_{\rm Tor}$ available with those for which we have $M_{\rm BH}$ we obtain a final sample of 549 objects, of which 289 are unobscured [$\log (N_{\rm H}/\rm cm^{-2})<22$] and 260 are obscured [$\log (N_{\rm H}/\rm cm^{-2})\geq 22$] AGN. For 19 of these objects, only an upper limit on $L_{\rm Tor}$ is available.

For all objects in our final sample, the intrinsic 14--150\,keV and 12\,$\mu$m luminosities were calculated using the redshifts and redshift-independent distances reported in \citet{Koss:2022bb}. We calculated the Eddington ratio $\lambda_{\rm Edd}$ from the Eddington luminosity: $L_{\rm Edd}=\frac{4\pi G M_{\rm BH} m_{\rm p}c}{\sigma_{\rm T}}$, where $G$ is the gravitational constant, $m_{\rm p}$ is the mass of the proton, $c$ is the speed of light, and $\sigma_{\rm T}$ is the Thomson cross-section. The bolometric luminosity was calculated by either adopting the $\lambda_{\rm Edd}$-dependent 2--10\,keV bolometric corrections of \citeauthor{Vasudevan:2007qt} [\citeyear{Vasudevan:2007qt}; $\kappa_{2-10}=\kappa(\lambda_{\rm Edd})$], or considering a 14--150\,keV bolometric correction of $\kappa_{14-150}=8.48$ ($L_{\rm Bol}=\kappa_{14-150}\times L_{14-150}$). The latter is equivalent to a 2--10\,keV bolometric correction of $\kappa_{2-10}=20$ \citep{Vasudevan:2007qt} for an X-ray photon index of $\Gamma=1.8$, consistent with the typical value of {\it Swift}/BAT AGN \citep{Ricci:2017cj}. The $\lambda_{\rm Edd}$-dependent 2--10\,keV bolometric corrections of \citeauthor{Vasudevan:2007qt}  were implemented as in \citet{Ricci:2017ss}, considering $\kappa_{2-10}= 20$ for $\lambda_{\rm Edd}\leq 0.04$, $\kappa_{2-10}=70$ for $\lambda_{\rm Edd} \geq 0.4$, and follow $\kappa_{2-10}\propto \lambda_{\rm Edd}^{0.54}$ over the range $0.04 < \lambda_{\rm Edd} < 0.4$. To also consider objects for which only an upper limit in $L_{\rm Tor}$ is available, all medians were calculated with the Kaplan-Meier estimator within the \textsc{asurv} package \citep{Feigelson:1985qv,Isobe:1986ys}, using a \textsc{python} implementation (see \S5 in \citealp{Shimizu:2016hc} for details). We also verified our results using other \textsc{python} implementation of the survival analysis approach, such as \textsc{scikit-survival} \citep{Polsterl:2020nh}.

\section{The covering factor of dust}\label{sect:CFdust}

While most studies carried out in the past decade have used 
\begin{equation}\label{eq:ratiolum}
R\equiv L_{\rm Tor}/L_{\rm Bol}
\end{equation} 
as a direct proxy of the covering factor of the circumnuclear dust (CF), \citet{Stalevski:2016kl} have shown that the covering factor of the torus is not directly proportional to $R$; instead, it is related by a more complex function due to the radiative transfer effects. In particular, \citet{Stalevski:2016kl} illustrated how the anisotropic emission of both the accretion disk and the dusty torus plays a strong role in the relation between $R$ and $CF$. The authors showed that $R$ overestimates high covering factors and underestimates low covering factors for unobscured AGN, while always underestimating the covering factors for obscured sources. \citet{Stalevski:2016kl} provide corrections to take all of these effects into account (see Table\,1 of their paper). These corrections do not yet take into account the effect of the SMBH spin, which could also change the radiation pattern of the accretion flow (e.g., \citealp{Campitiello:2018cz,Ishibashi:2019jy}). Here we will focus on the dust covering factors obtained by applying the corrections suggested by \citet{Stalevski:2016kl}. These corrections are given in the form of polynomials of $R$, and depend on the optical depth of the silicate feature ($\tau_{9.7}$), and whether the disk and the torus are misaligned. For pole-on (i.e., unobscured) objects, the covering factor of the dust is given by

 \begin{figure}
\centering
 %% 1st image
 %% 2nd image
\includegraphics[width=0.48\textwidth]{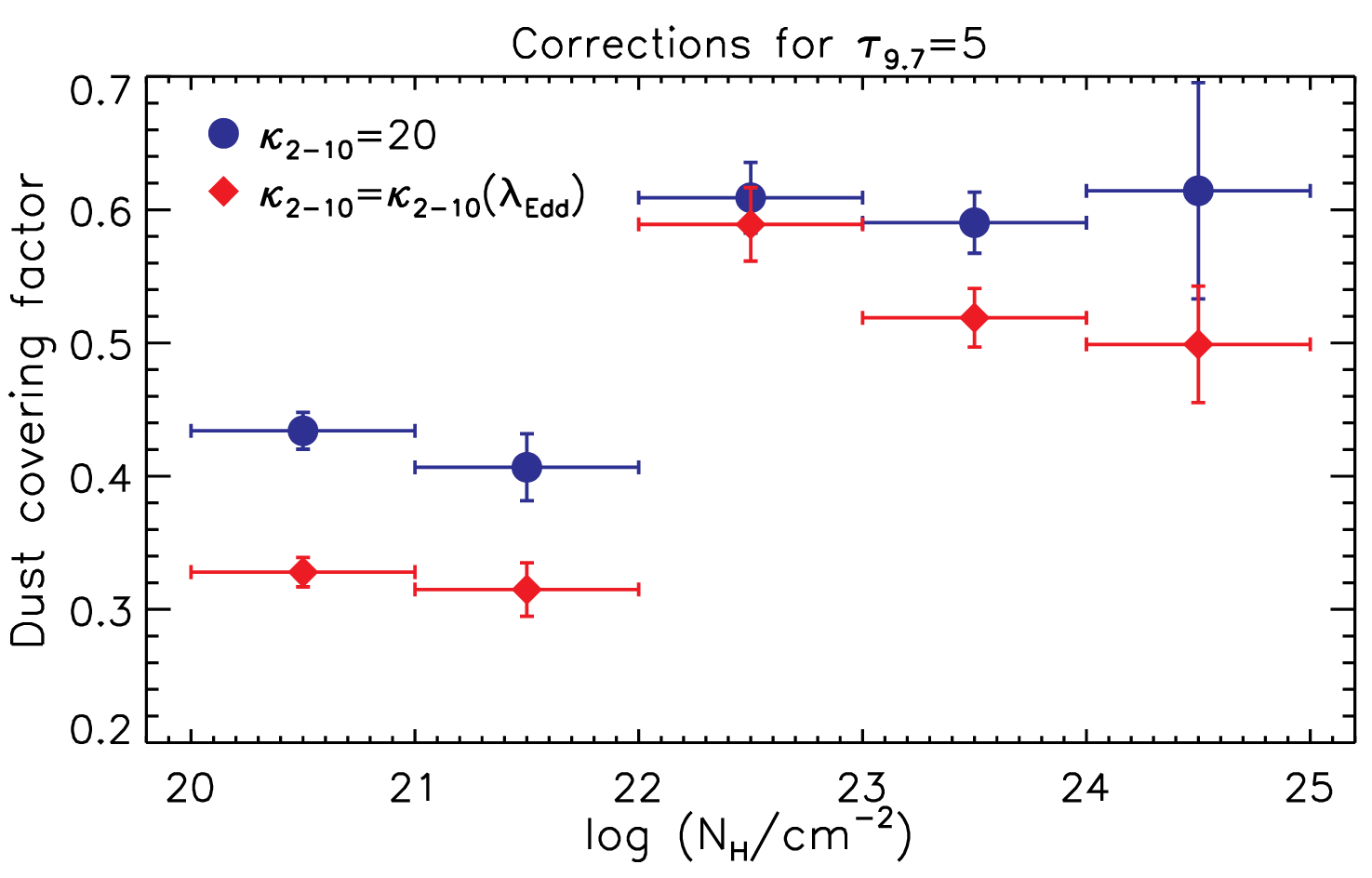}
% %% caption
% \begin{minipage}[t]{1\textwidth}
  \caption{Median dust covering factor (see \S\ref{sect:CFdust} and Eqs.\ref{eq:CFunobs}-\ref{eq:CFobs}) versus column density for the sources of our sample. The bolometric luminosity was estimated by using a fixed 2--10\,keV bolometric corrections ($\kappa_{2-10}=20$, blue circles) or by using the Eddington ratio dependent corrections of \citeauthor{Vasudevan:2007qt} (\citeyear{Vasudevan:2007qt}, red diamonds). The dust covering factors were calculated assuming an optical depth at 9.7$\mu$m of $\tau_{9.7}=5$ and Eqs.\,\ref{eq:CFunobs} and \ref{eq:CFobs}.}
\label{fig:cfdust_vsNH}
% \end{minipage}
\end{figure}

\begin{equation}\label{eq:CFunobs}
CF^{\rm unobs}=a_{4}R^4+a_{3}R^3+a_{2}R^2+a_{1}R+a_{0},
\end{equation}

while for edge-on (i.e., obscured) AGN

\begin{equation}\label{eq:CFobs}
CF^{\rm obs}=b_{3}R^3+b_{2}R^2+b_{1}R+b_{0}
\end{equation}

We considered here $\tau_{9.7}=5$ and, following the prescriptions of \citet{Stalevski:2016kl} we set the maximum value of the ratio between the torus and bolometric AGN luminosity ($R_{\rm max}$) to 1.3 for unobscured objects and $R_{\rm max}=1$ for obscured AGN. In the following we will use the dust covering factors obtained from Eqs.\ref{eq:CFunobs} and \ref{eq:CFobs}. We also tested the scenario in which the torus and the disk are misaligned (with $\tau_{9.7}=5$), in which case $R_{\rm max}$ for unobscured objects was set to 1.5 and 1.6, for an inclination of the disk with respect to the torus of $15^{\circ}$ and $30^{\circ}$, respectively. We found that this does not significantly affect any of the results reported here.

\begin{figure*}
\centering
 \includegraphics[width=0.48\textwidth]{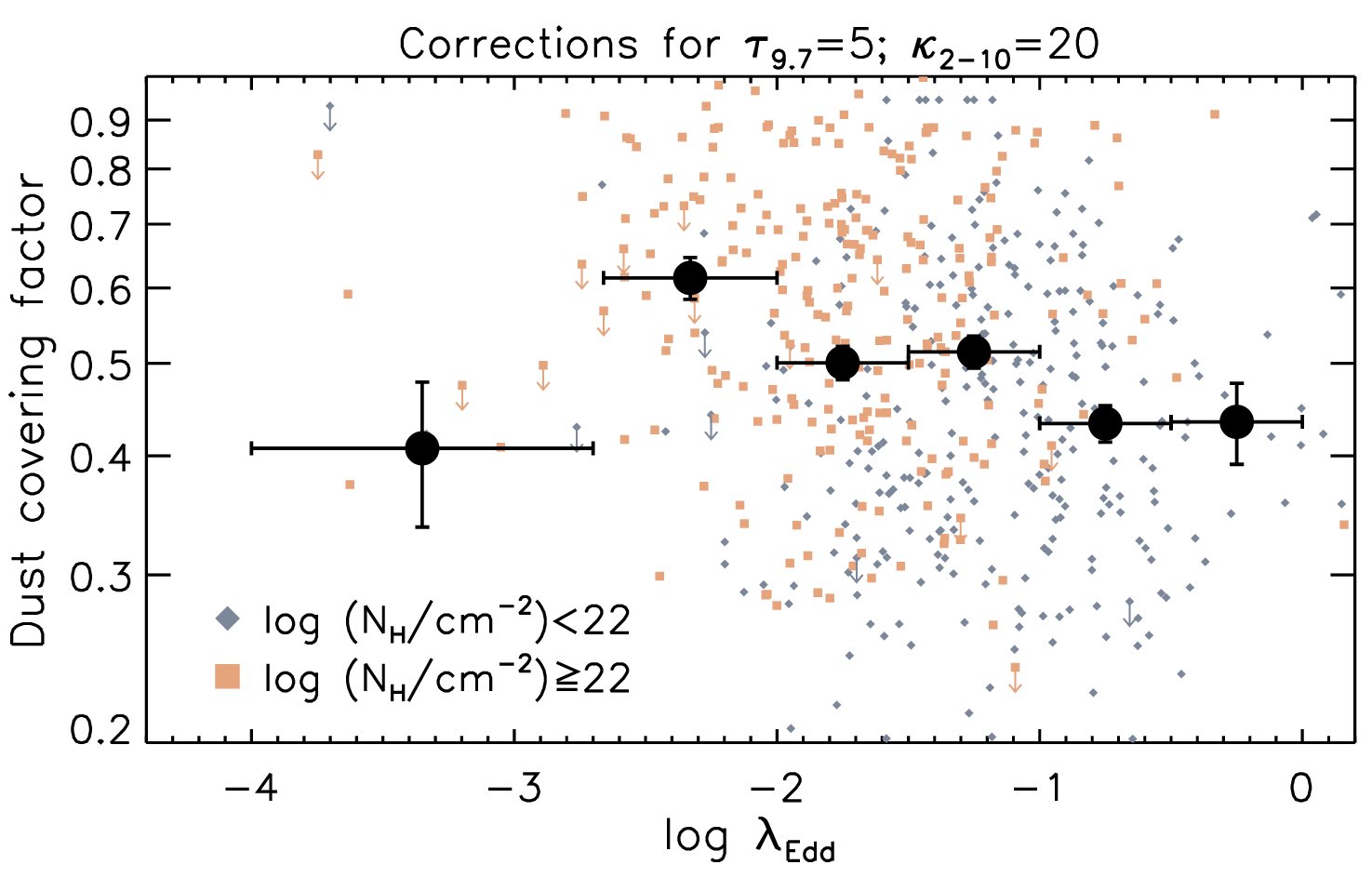}
 \includegraphics[width=0.48\textwidth]{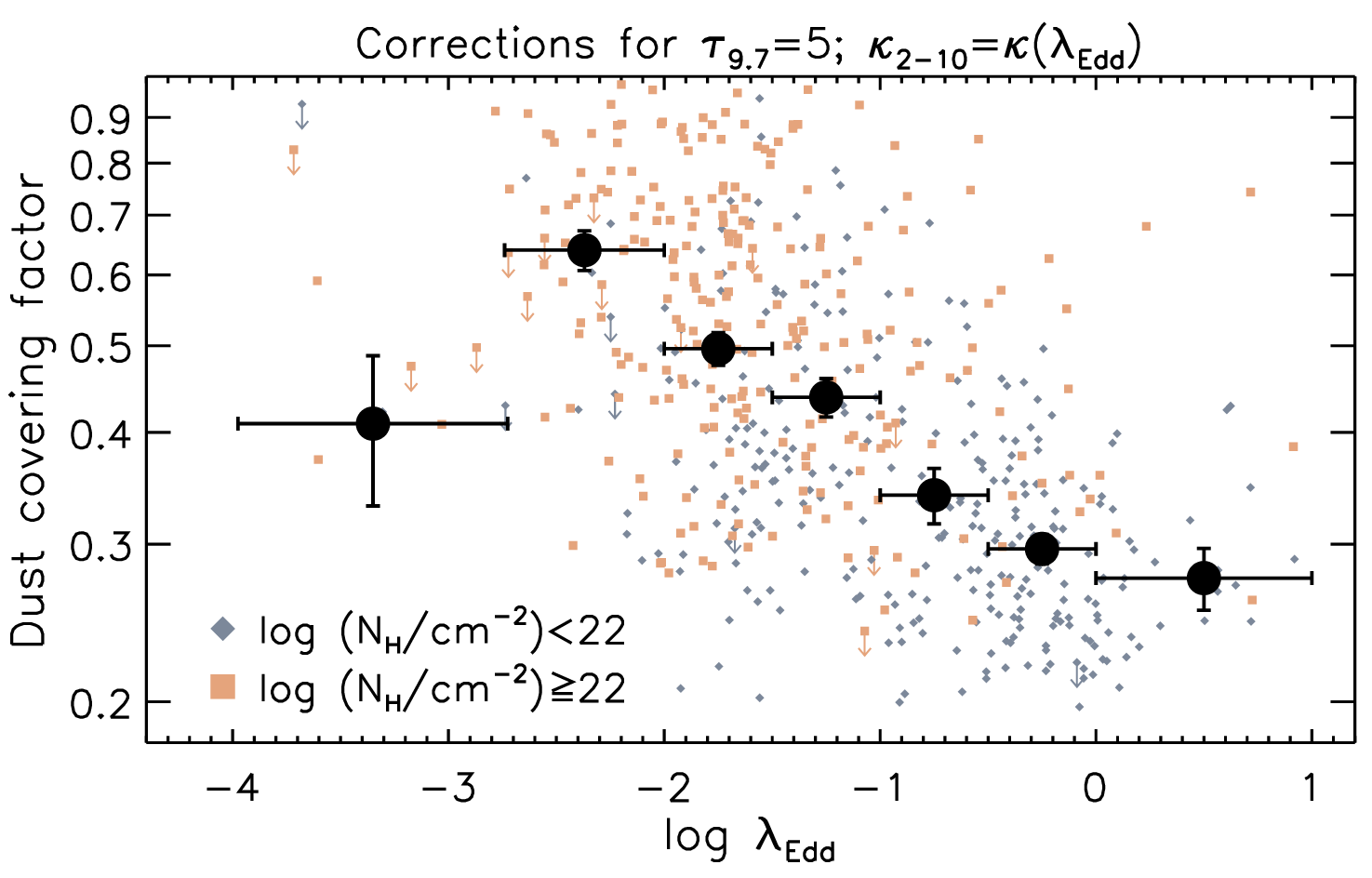}
\par\bigskip
\includegraphics[width=0.48\textwidth]{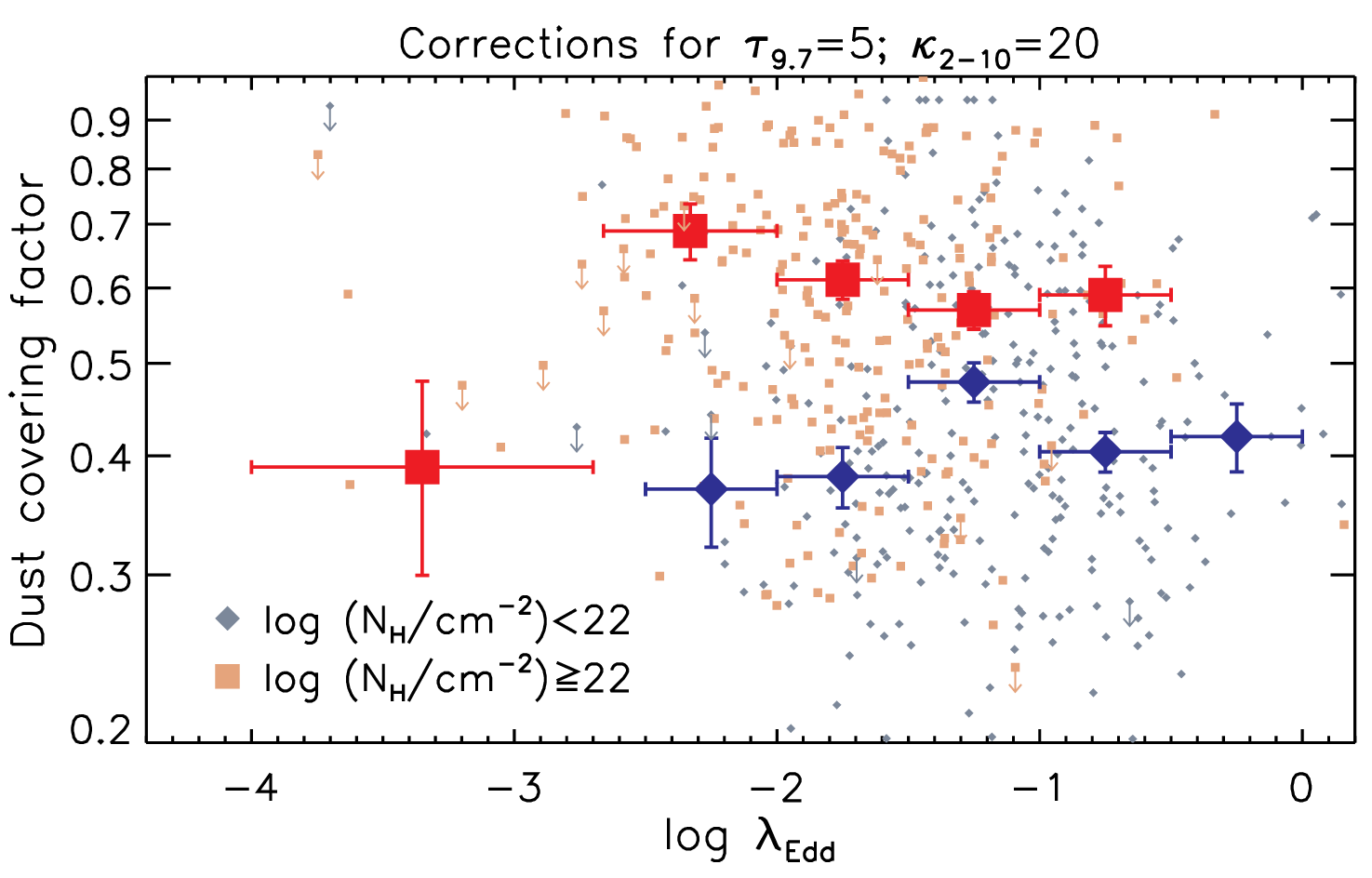}
\includegraphics[width=0.48\textwidth]{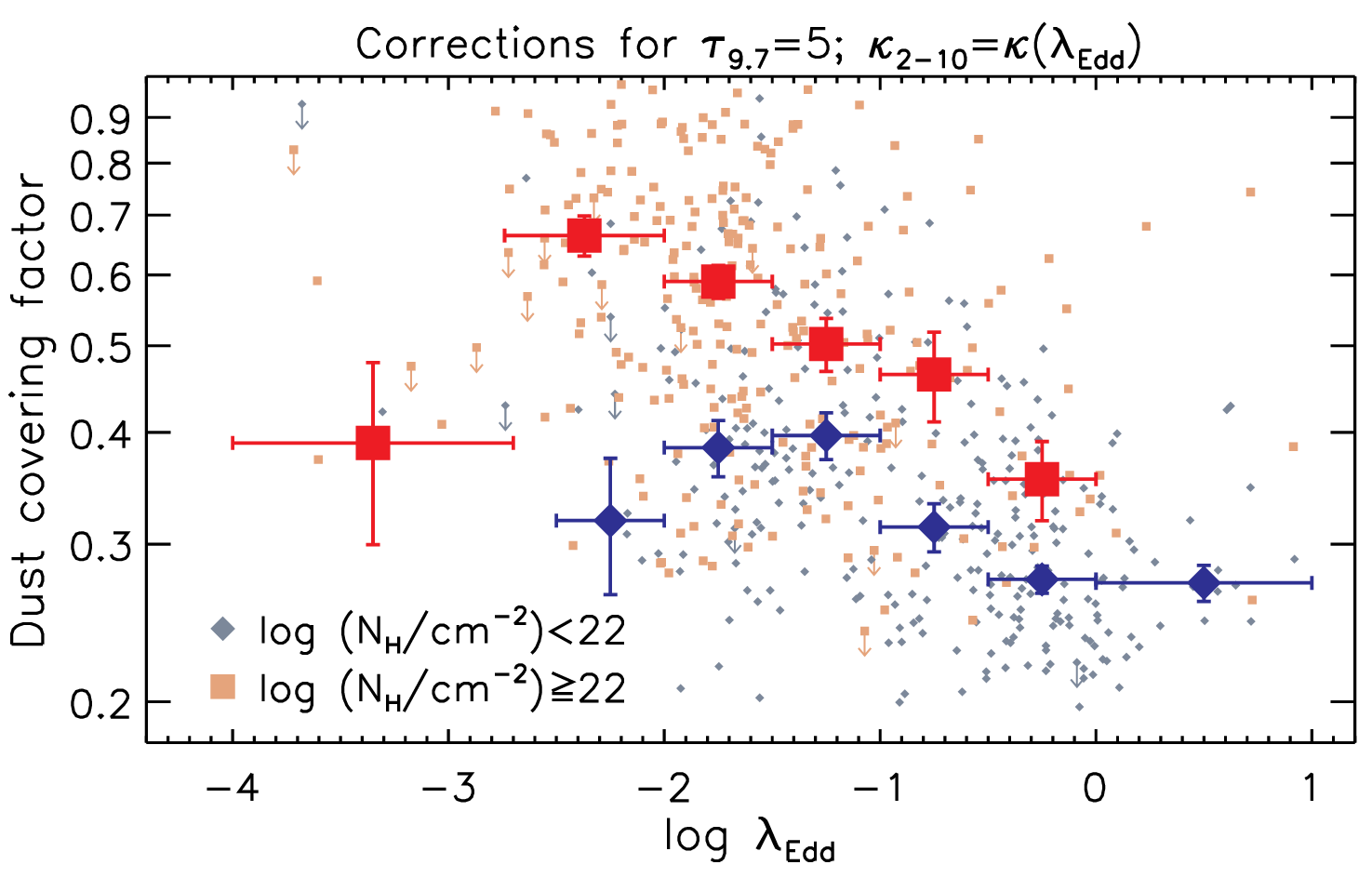}

%% 1st image
 %% 2nd image
% %% caption
% \begin{minipage}[t]{1\textwidth}
  \caption{Dust covering factor (see \S\ref{sect:CFdust} and Eqs.\ref{eq:CFunobs}-\ref{eq:CFobs}) versus Eddington ratio for the sources of our sample. The large points in the plots show the median values for the whole population (black circles; top panels), for unobscured and obscured AGN (blue diamonds and red squares, respectively; bottom panels). The left panels show the case in which the bolometric luminosity is estimated by using a fixed 2--10\,keV bolometric correction ($\kappa_{2-10}=20$), while in the right panels the bolometric luminosity was estimated by using the Eddington ratio-dependent corrections of \citet{Vasudevan:2007qt}. The dust covering factors were calculated assuming an optical depth at 9.7$\mu$m of $\tau_{9.7}=5$ and Eqs.\,\ref{eq:CFunobs} and \ref{eq:CFobs}.}
\label{fig:CFdustvsEddratio_corr_taufive}
% \end{minipage}
\end{figure*}

The median dust covering factors obtained using the approach described above are $0.50\pm0.01$ and $0.41\pm0.02$ for $\kappa_{2-10}=20$ and $\kappa_{2-10}=\kappa(\lambda_{\rm Edd})$, respectively. Interestingly, as observed in \cite{Ichikawa:2019zz}, these values are significantly lower than the covering factor obtained by X-ray observations for the same sample ($\sim 70\pm2\%$, \citealp{Ricci:2015tg,Ricci:2017ss}). This difference could be associated with obscuration from the broad line region (e.g., \citealp{Davies:2015ve,Ichikawa:2019zz}) or from dust-free gas associated with outflows produced by the AGN. In Fig.\,\ref{fig:cfdust_vsNH} we show the dust covering factor of BASS AGN versus the column density inferred by X-ray observations by considering fixed bolometric corrections (blue circles) or Eddington ratio-dependent bolometric corrections (red diamonds). In both cases, the dust covering factor of the obscuring material in an unobscured AGN is $\sim 1.5-2$ times lower than for an obscured AGN. This is in agreement with previous studies carried out by applying a torus model to the IR emission of nearby AGN (e.g., \citealp{Ramos-Almeida:2011eb}). As proposed by \citet{Elitzur:2012dq}, this could be explained simply by the fact that AGN in which the covering factor of the obscuring material is large are more likely to be observed as obscured/type-2, due to the larger fraction of obscured lines-of-sight.

\begin{figure*}
\centering
 %% 1st image
 %% 2nd image
\includegraphics[width=0.748\textwidth]{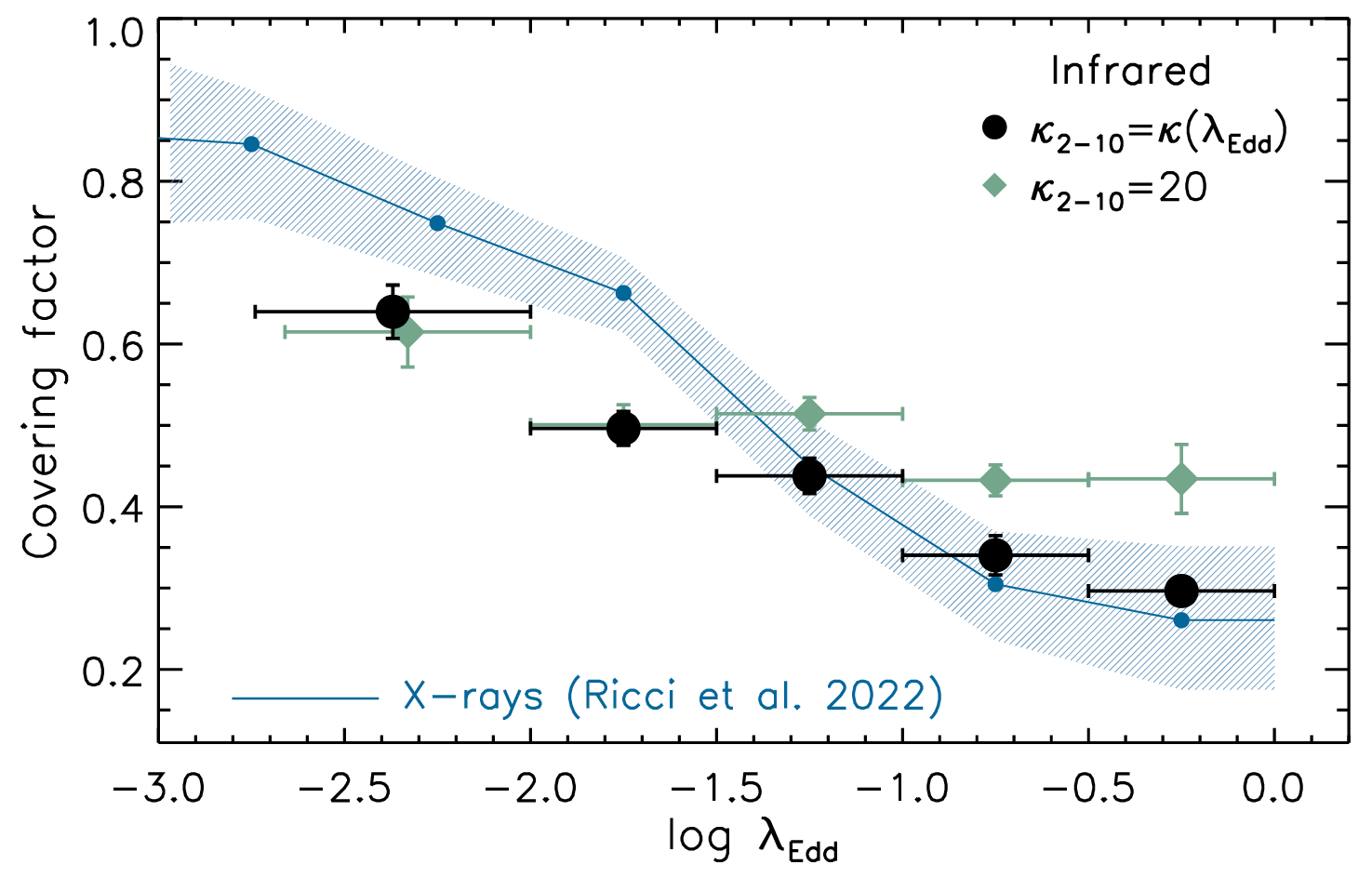}
% %% caption
% \begin{minipage}[t]{1\textwidth}
  \caption{Covering factor of dust versus Eddington ratio for the sources of our samples considering the Eddington ratio dependent corrections of \citet{Vasudevan:2007qt} (black circles) and fixed 2--10\,keV bolometric corrections ($\kappa_{2-10}=20$; green diamonds). The solid blue line shows the covering factor of gas inferred from the fraction of obscured sources (\citealp{Ricci:2022xp}, assuming $\kappa_{2-10}=20$, see \citealp{Ricci:2017ss} for a discussion on how bolometric corrections affect the $f_{
 \rm obs}-\lambda_{\rm Edd}$ trend), while the hatched area represents the 1$\sigma$ uncertainty. The dust covering factors were calculated assuming an optical depth at 9.7$\mu$m of $\tau_{9.7}=5$ and Eqs.\,\ref{eq:CFunobs} and \ref{eq:CFobs}.}
\label{fig:CFdust_IRvsXray}
% \end{minipage}
\end{figure*}

\subsection{The dependence of the covering factor on the Eddington ratio}\label{sect:cfvslumeddratio}

In Fig.\,\ref{fig:CFdustvsEddratio_corr_taufive} we illustrate the dust covering factor versus the Eddington ratio for the sources of our sample. This was done considering $\kappa_{2-10}=20$ (left panels) and $\kappa_{2-10}=\kappa(\lambda_{\rm Edd})$ (right panels). In the top panels we show the median values of the dust covering factor for the whole sample (large black circles), while in the bottom panels we illustrate the median values for obscured (large red squares) and unobscured (large blue diamonds) AGN. The medians were calculated considering the upper limits, as outlined in \S\ref{sect:sampledata}. In both cases, we find a decrease of the covering factor with the Eddington ratio for AGN with $\log\lambda_{\rm Edd}>-3$, which is steeper when considering the $\lambda_{\rm Edd}$-dependent bolometric corrections of \citet{Vasudevan:2007qt}. The decrease starts at $\lambda_{\rm Edd}\sim 10^{-2}$, similar to what was observed for the covering factor inferred from the fraction of obscured sources \citep{Ricci:2017ss,Ricci:2022xp}. The bottom figures show that obscured AGN tend to have larger covering factors than unobscured AGN, as discussed in \S\ref{sect:CFdust}, regardless of their Eddington ratio. When assuming a fixed bolometric correction ($\kappa_{2-10}=20$), the relation between the dust covering factor and $\lambda_{\rm Edd}$ for all AGN is mostly driven by the difference in covering factor between obscured and unobscured AGN, and the fact that unobscured AGN are mostly found at higher Eddington ratios \citep{Ricci:2017ss}. On the other hand, when assuming $\lambda_{\rm Edd}$-dependent bolometric corrections, one can observe a clear decrease of the dust covering factor with $\lambda_{\rm Edd}$ also for obscured and unobscured AGN. It should be noted that there is an inherent dependence on the choice of bolometric correction in the values of the covering factor, since they are a function of the ratio between the IR AGN luminosity and the bolometric luminosity. However, a recent study focused on a sample of $\sim 250$ unobscured BASS AGN with simultaneous X-ray and optical/UV observations has shown that the Eddington ratio is the main driver of the bolometric corrections (Gupta et al. in prep.). To ensure the robustness of our results we tested two additional bolometric corrections: the luminosity-dependent bolometric corrections from \citet{Lusso:2012it} and the more recent formulation of the Eddington-ratio dependent bolometric corrections from Gupta et al. (in prep.). In both cases we recover trends very similar to those presented here by assuming the bolometric corrections of \citet{Vasudevan:2007qt}, with two differences: i) a flatter decrease with $\lambda_{\rm Edd}$ for the obscured AGN population when considering the bolometric corrections from \citet{Lusso:2012it}; ii) slightly higher covering factors when using the Gupta et al. (in prep.) bolometric corrections ($\Delta CF\sim 0.1$). Similar to what we found using $f_{\rm obs}$ \citep{Ricci:2017ss}, and regardless of the choice of the bolometric correction, we see a decrease in the covering factor at low Eddington ratios, which could be associated with an evolutionary sequence of AGN in the $N_{\rm H}-\lambda_{\rm Edd}$ plane (\citealp{Ricci:2022xp}; see \S\ref{sect:RRgrowth}).

In \citet{Ricci:2017ss,Ricci:2022xp}, we have shown that the main driver of the covering factor of the Compton-thin obscuring material is the Eddington ratio, and that above the effective Eddington limit for dusty gas ($\lambda_{\rm Edd}^{\rm eff}$) with $N_{\rm H}\simeq 10^{22}\rm\,cm^{-2}$ [$\lambda_{\rm Edd}^{\rm eff}(10^{22}\rm\,cm^{-2})\simeq 0.02$, see \citealp{Fabian:2006lq,Fabian:2008hc,Fabian:2009ez,Arakawa:2022tl}] most AGN are unobscured. The effective Eddington limit is given by the ratio between the Thomson cross section ($\sigma_{\rm T}$) and the effective cross section of the interaction ($\sigma_{\rm i}$), i.e. $\lambda_{\rm Edd}^{\rm eff}= \sigma_{\rm T}/\sigma_{\rm i}$. The value of $\sigma_{\rm i}$ depends on the physical properties of the material and is typically $\sigma_{\rm i} \gg \sigma_{\rm T}$, which implies that $\lambda_{\rm Edd}^{\rm eff}<1$. For $\lambda_{\rm Edd}\geq \lambda_{\rm Edd}^{\rm eff}(N_{\rm H})$ the obscuring material is expected to be expelled by radiation pressure, and dusty outflows would be expected to populate the polar region of the system (see \S\ref{sec:polardust}). Results consistent with those reported in \citet{Ricci:2017ss} were also obtained using the ratio between the Eddington ratio distribution functions of obscured and unobscured {\it Swift}/BAT AGN \citep{Ananna:2022qc,Ananna:2022tr}, and by using the results of broad-band X-ray spectroscopy (e.g., \citealp{Zhao:2020ql}, see also \citealp{Ogawa:2021um}) after fixing the inclination angle with respect to the torus to the value obtained from careful modeling of the narrow-line region \citep{Fischer:2013hs}. 

In Fig.\,\ref{fig:CFdust_IRvsXray} we illustrate the dust covering factors versus $\lambda_{\rm Edd}$ obtained in the infrared (black circles and green diamonds) together with the covering factor obtained in the X-rays by the recent study of \citet{Ricci:2022xp} (blue line). There is generally a rather good agreement between the values obtained in the IR (for dusty gas) and in the X-rays (for dust-free and dusty gas). The slightly less steep trend observed in the IR might be ascribed to the fact that as one moves toward higher $\lambda_{\rm Edd}$ the dust expelled by radiation pressure would still contribute to the IR emission in the form of a polar component (see also \citealp{Garcia-Bernete:2019mp,Toba:2021ys}). The relation between the covering factor and $\lambda_{\rm Edd}$ implies that, in addition to the inclination angle ($\theta_{\rm\,i}$), the observational classification of AGN into obscured and unobscured is driven by the Eddington ratio, in what was defined as the {\it radiation-regulated unification model} \citep{Ricci:2017ss}. If the filling factor of the obscuring clouds is large, the half-opening angle of the torus for a given value of $\lambda_{\rm Edd}$ can be deduced from $\theta_{\rm\,OA}(\lambda_{\rm Edd})=\cos^{-1}[f_{\rm\,obs}^{\,*}(\lambda_{\rm Edd})]$, where $f_{\rm\,obs}^{\,*}$ is the fraction of all obscured sources (i.e. both Compton-thin and Compton-thick). In this scheme, a source will be unobscured if the inclination angle is smaller than the half-opening angle of the torus [$\theta_{\rm\,i}<\theta_{\rm\,OA}(\lambda_{\rm Edd})$], while it will be obscured if $\theta_{\rm\,i}> \theta_{\rm\,OA}(\lambda_{\rm Edd})$. 

\subsection{Dusty outflows and polar dust}\label{sec:polardust}

\begin{figure*}
 \centering
 \includegraphics[width=0.7\textwidth]{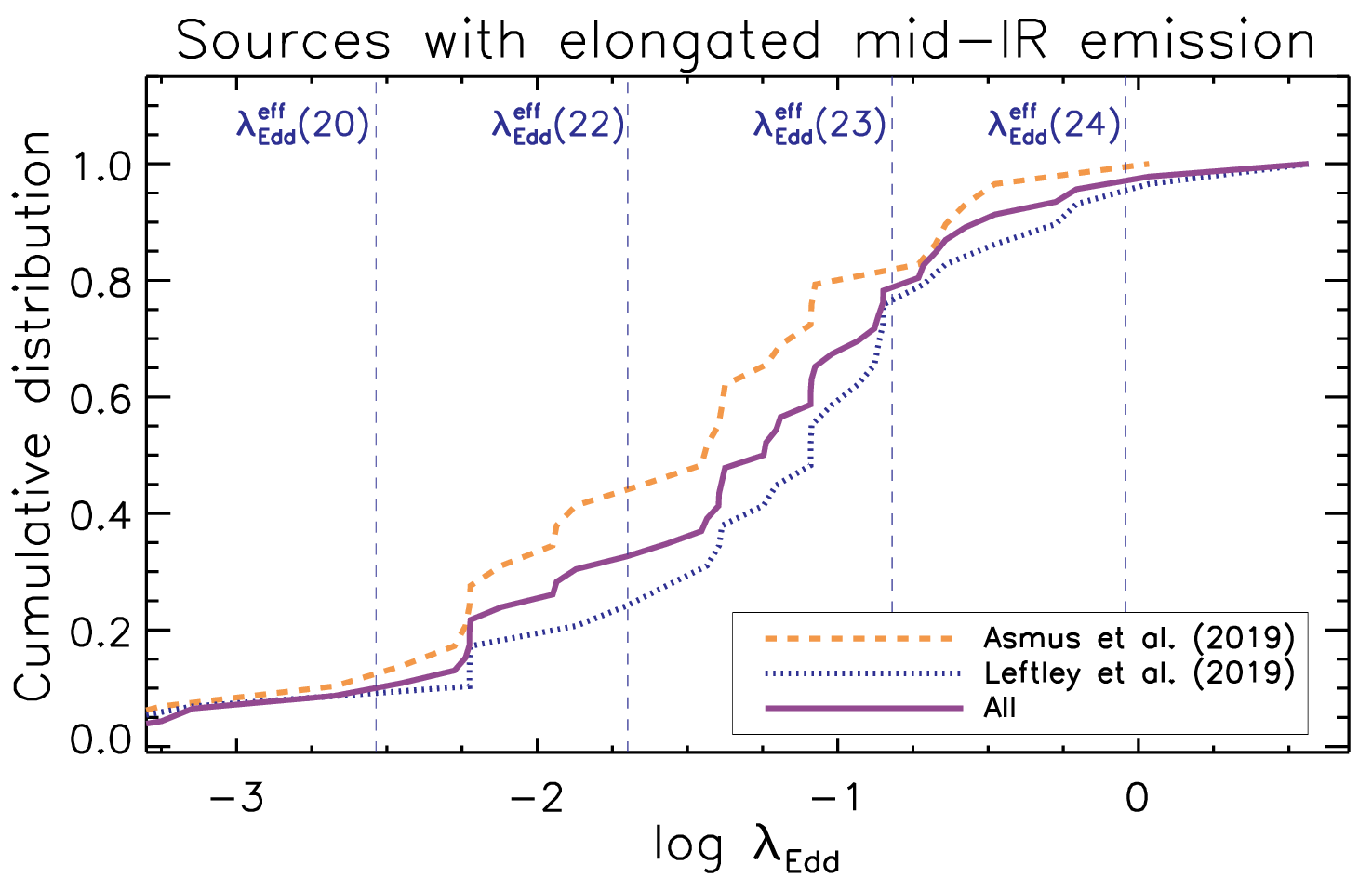}
 \caption{Cumulative Eddington ratio distribution of sources showing polar mid-IR emission. Cumulative distribution of $\lambda_{\rm Edd}$ for sources showing elongated mid-IR emission from interferometric (\citealp{Leftley:2019zj}, dotted blue line) and high-resolution non-interferometric observations (\citealp{Asmus:2019wz}, dashed orange line). The continuous line shows the total cumulative distribution. The dashed vertical lines represent the effective Eddington limit for dusty gas with $N_{\rm H}=10^{20}\rm\,cm^{-2}$ [$\lambda_{\rm Edd}^{\rm\,eff}(20)$], $N_{\rm H}=10^{22}\rm\,cm^{-2}$ [$\lambda_{\rm Edd}^{\rm\,eff}(22)$], $N_{\rm H}=10^{23}\rm\,cm^{-2}$ [$\lambda_{\rm Edd}^{\rm\,eff}(23)$] and $N_{\rm H}=10^{24}\rm\,cm^{-2}$ [$\lambda_{\rm Edd}^{\rm\,eff}(24)$] for dust grain abundance consistent with the ISM.}
\label{fig:cdf_polar}
\end{figure*}

Over the past few years, interferometric studies \citep{Jaffe:2004fj,Wittkowski:2004qv,Tristram:2007kq,Raban:2009bx,Honig:2012hq,Honig:2013wf,Burtscher:2013xu,Lopez-Gonzaga:2014rm,Tristram:2014ss,Lopez-Gonzaga:2016cj,Isbell:2022gz,Gamez-Rosas:2022gp} and high-spatial resolution observations carried out by 8-10\,m telescopes \citep{Asmus:2016uf,Garcia-Bernete:2016sg} have shown that a significant fraction of the mid-IR emission of AGN is elongated in the polar direction, and models including a disk and a wind can reproduce the near- to mid-IR properties of local AGN \citep{Honig:2017qp}. The elongated mid-IR emission, which is thought to originate in a hollow cone (e.g., \citealp{Stalevski:2017qv,Stalevski:2019gl}), could be related to dusty outflows arising at $\lambda_{\rm Edd}\geq \lambda_{\rm Edd}^{\rm\,eff}(N_{\rm H})$ (e.g., \citealp{Honig:2012hq,Venanzi:2020dd,Tazaki:2020js}). Such outflows could also be associated to larger-scale winds observed in AGN (e.g., \citealp{Kakkad:2016qv,Rojas:2020gz,Stacey:2022cv,Musiimenta:2023qj}).

Consistent with the idea that radiation pressure is responsible for the extended IR emission in AGN, the Eddington ratio of the objects currently known to show polar mid-IR emission is typically rather high. Some of the sources where polar dust was observed, such as Circinus \citep{Tristram:2014ss}, NGC\,424 \citep{Honig:2012hq}, NGC\,1068 \citep{Lopez-Gonzaga:2014rm} are among the obscured AGN with the highest Eddington ratio in our sample (see also \citealp{Leftley:2019zj}), and several of them lie in the forbidden region of the $N_{\rm H}-\lambda_{\rm Edd}$ diagram. Indeed, \cite{Alonso-Herrero:2021id} found that AGN with polar mid-IR (MIR) emission show intermediate column densities [$\log (N_{\rm H}/\rm cm^{-2}) \sim 22.5-23$] and Eddington ratios ($\log \lambda_{\rm Edd} \sim -1.75$ to $-1$). \cite{Garcia-Bernete:2022pk} fitted the IR spectra of nearby AGN from the {\it Swift}/BAT sample and found that AGN that are best reproduced by models that include a polar dust component typically have $\log \lambda_{\rm Edd} \gtrsim -2.5$ (see also \citealp{Gonzalez-Martin:2019gy}). \cite{Yamada:2023cg} showed, for a sample of nearby AGN undergoing mergers that the contribution of the polar component to the IR AGN luminosity increases for $\log \lambda_{\rm Edd}\gtrsim -3$ (see their Fig.\,21). However, it should be noted that the forbidden region is defined for the line-of-sight column density $N_{\rm H}$, so even sources accreting at $\lambda_{\rm Edd}< \lambda_{\rm Edd}^{\rm\,eff}(N_{\rm H})$ might be in the process of expelling some material if, for some inclination angles, the clouds surrounding the SMBH have column densities [$N_{\rm H}(\theta_{\rm\,i})$] lower than the value inferred for the line-of-sight, and such that $\lambda_{\rm Edd}\geq \lambda_{\rm Edd}^{\rm\,eff}[N_{\rm H}(\theta_{\rm\,i})]$. 

In Figure\,\ref{fig:cdf_polar} we illustrate the cumulative Eddington ratio distribution of sources showing mid-IR polar emission from interferometric (\citealp{Leftley:2019zj}, dotted blue line) and high-resolution single dish observations (\citealp{Asmus:2019wz}, dashed orange line), most of which are part of our {\it Swift}/BAT-selected sample. Interestingly, all of the sources showing elongated mid-IR emission are accreting above the Eddington limit for dusty gas with $N_{\rm H}=10^{20}\rm\,cm^{-2}$, and more than half of them ($61\pm10\%$) at $\lambda_{\rm Edd}\geq \lambda_{\rm Edd}^{\rm\,eff}(10^{22}\rm\,cm^{-2})$. A possible caveat of these studies is that we might preferentially detect extended IR emission in AGN with higher luminosities and Eddington ratios because of the stronger contrasts with respect to the host galaxy, whose emission could hide extended IR emission at low accretion rates.

\begin{figure*}
\centering
 %% 1st image
 %% 2nd image
\includegraphics[width=0.748\textwidth]{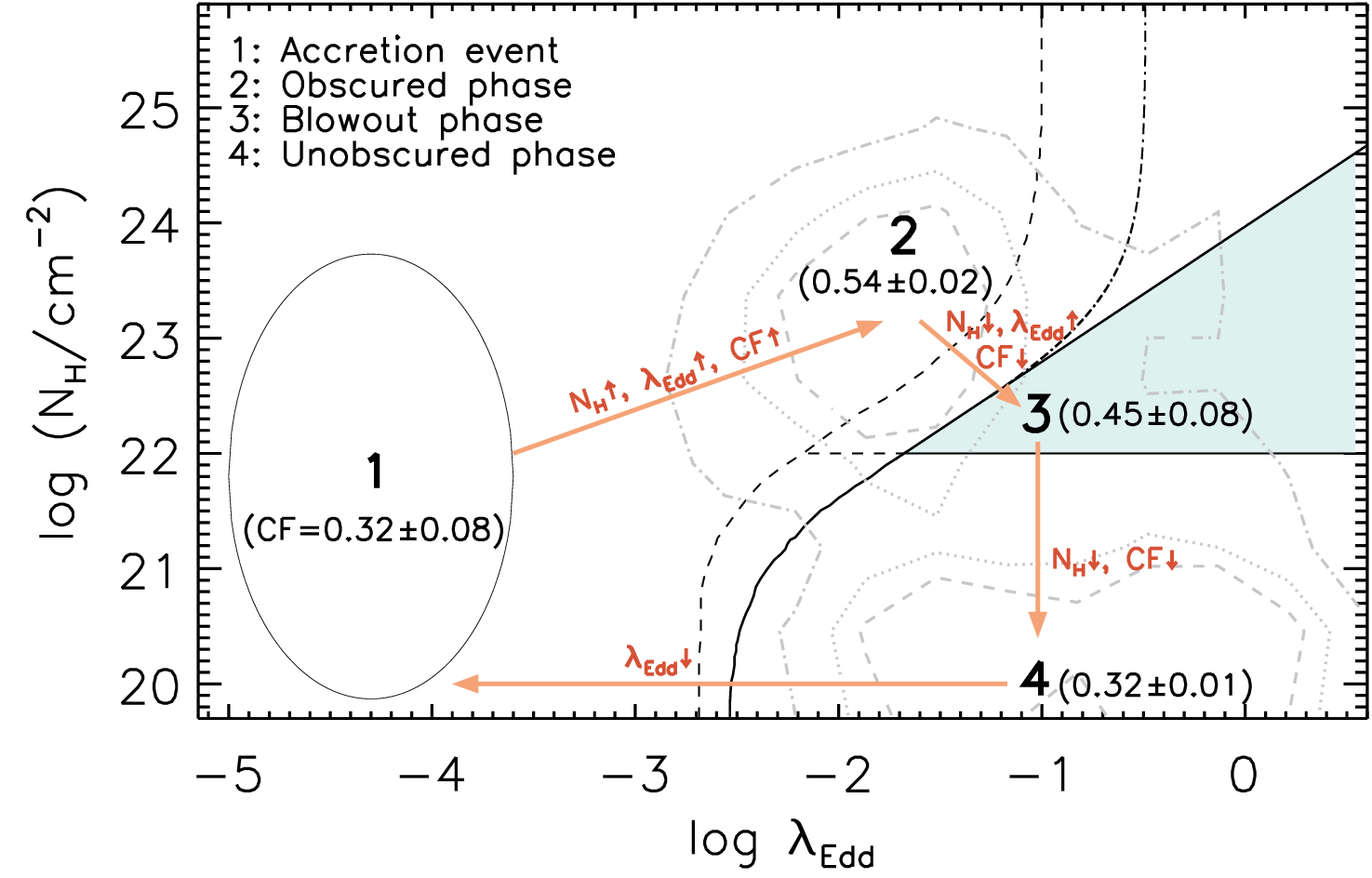}
% %% caption
% \begin{minipage}[t]{1\textwidth}
  \caption{Schematic of radiation-regulated growth of SMBHs together with the median covering factors of dusty gas at different stages (see \S\ref{sect:RRgrowth}). The effective Eddington limit for dusty gas reported in \citet{Fabian:2009ez} is shown with the black solid lines, while the blowout region is the green area. The plot also shows the effective Eddington limit when including infrared radiation trapping \citep{Ishibashi:2018it}, adapted to the values of \citet{Fabian:2009ez}, similarly to what was done by \citeauthor{Lansbury:2020ee} (\citeyear{Lansbury:2020ee}; dashed-dotted lines). The effective Eddington limit for dusty gas reported by \citet{Ishibashi:2018it} is shown by the dashed line. The assumed maximum contribution to $N_{\rm H}$ of gas from the host galaxy is illustrated by the horizontal line at $\log (N_{\rm H}/\rm cm^{-2})=22$. The number density contours in the $-3 \leq \log \lambda_{\rm Edd} \leq 1$ range are shown by the grey dashed (50\%), dotted (68\%), and dotted-dashed (90\%) lines. The bolometric corrections of \citet{Vasudevan:2007qt} were used both for CF and $\lambda_{\rm Edd}$. }
\label{fig:RRgrowth_CF}
% \end{minipage}
\end{figure*}

\subsection{Radiation-regulated SMBH growth}\label{sect:RRgrowth}

As argued by \citet{Ricci:2017ss}, the radiation-regulated unification model could be dynamic, with AGN moving in the $N_{\rm H}-\lambda$ diagram during their lifetime (see also \citealp{Jun:2021sq,Toba:2022hq}). \citet{Ricci:2022xp} showed that the breaks observed in both the Eddington ratio distribution function and the luminosity function \citep{Ananna:2022qc} correspond to the $\lambda_{\rm Edd}$ and $L_{14-150\rm\,keV}$ at which AGN transition from having most of their sky covered by absorbing material to having most of their sky devoid of obscuring material (see Fig.\,4 of . \citealp{Ricci:2022xp} ). This implies that the majority of the SMBH growth at $z\sim0$ happens when most of the AGN is covered by gas and dust, and that AGN accreting above the Eddington limit for dusty gas are rarer. \citet{Ricci:2022xp} proposed that this could be related to the lower amount of material available to feed the SMBH when the AGN is accreting very rapidly, due to the depletion of the gas reservoir caused by radiation pressure, and that AGN could move in the $N_{\rm H}-\lambda_{\rm Edd}$ plane during their life cycle (see Fig.\,\ref{fig:RRgrowth_CF}). In this picture SMBHs start their growth phase at low $\lambda_{\rm Edd}$ in a mostly unobscured phase (stage 1). As the fuel moves toward the supermassive black hole, $\lambda_{\rm Edd}$, $N_{\rm H}$ and the covering factor would gradually increase, leading the source to be preferentially observed as obscured by observers with random inclination angles (stage\,2). Once $\lambda_{\rm Edd}\gtrsim \lambda_{\rm Edd}^{\rm eff}$ the AGN would start expelling the obscuring material (stage\,3), which would lead to a decrease in both its covering factor and typical $N_{\rm H}$, and the source would be preferentially observed as an unobscured AGN (stage\,4). This last phase is expected to be rather short, and to last $\sim 3-30$\,Myr (\citealp{Ananna:2022tr}; assuming an AGN lifetime of $\sim 10^7-10^8$\,yr). Once the AGN accretes the rest of the material, it will move towards low Eddington ratios again (stage\,1). Interestingly, studying ALMA observations of 19 nearby AGN, \cite{Garcia-Burillo:2021qc} recently found a tentative decline of the average molecular gas surface density and H$_2$ column density for increasing Eddington ratios. They interpreted this, and the presence of molecular deficits on nuclear scales, as a signature of AGN feedback, and possibly an evolutionary sequence of AGN for increasing luminosities and Eddington ratios, similar to that proposed by  \citet{Ricci:2017ss} and \citet{Ricci:2022xp}.

Using the dust covering factor we estimated from IR luminosities, we can test this model and, in particular, check whether the typical covering factors change from stage\,1 to stage\,4. To do this, we used the bolometric corrections of \citealp{Vasudevan:2007qt}, and divided the $N_{\rm H}-\lambda$ into different regions. For stage\,1, we considered all objects with $\log \lambda_{\rm Edd} \leq -3.5$, while for stage\,2, we used all obscured AGNs with $-3.5 \leq \log \lambda_{\rm Edd} \leq \log \lambda_{\rm Edd}^{\rm eff}(N_{\rm H})$, using the original definition of $\lambda_{\rm Edd}^{\rm eff}(N_{\rm H})$ \citeauthor{Fabian:2009ez} (\citeyear{{Fabian:2009ez}}; black solid curve in Fig.\,\ref{fig:RRgrowth_CF}). For stage\,3 we considered all obscured AGNs in the blowout region (green area in Fig.\,\ref{fig:RRgrowth_CF}): $\log \lambda_{\rm Edd} \geq \log \lambda_{\rm Edd}^{\rm eff}(N_{\rm H})$. Finally, for stage\,4 we used all unobscured AGNs with  $\log \lambda_{\rm Edd} > -1.75$ [i.e. a value corresponding to $\sim \log \lambda_{\rm Edd}^{\rm eff}(10^{22}\rm\,cm^{-2})$]. The median covering factors we obtain agree well with the expected general trends: it is the lowest in stage\,1 ($0.32\pm0.08$), it increases by a factor two in stage\,2 ($0.54\pm0.02$), for then decreasing, likely due to the effect of radiation pressure, in stage\,3 ($0.45\pm0.08$) and stage\,4 ($0.32\pm0.01$). It should be stressed that in stage\,1 the covering factors could actually be lower than what was inferred here due to our AGN selection, which is missing AGN that accrete at very low Eddington ratios ($\log \lambda_{\rm Edd} < -4$). In fact, besides showing little absorption (see also \citealp{She:2018sg}), AGN accreting at very low $\lambda_{\rm Edd}$ also typically display faint reflection features (e.g., \citealp{Ptak:2004hw,Bianchi:2017oj,Diaz:2020ks,Diaz:2023od,Jana:2023qc}, see \citealp{Ho:2008et} for a review of the subject).

\section{Summary and conclusion}\label{sect:summary}
In this paper we studied a sample of 549 nearby non-blazar hard X-ray selected AGN (\S\ref{sect:sampledata}) to investigate the relation of the covering factor of dust around SMBHs and the Eddington ratio. We used the ratio of IR and bolometric AGN luminosity as a proxy of the covering factor of the circumnuclear dusty gas, considering the corrections of \citeauthor{Stalevski:2016kl} (\citeyear{Stalevski:2016kl}; \S\ref{sect:CFdust}). In the following, we summarize our main findings.
\begin{itemize}
\item We find that obscured AGN typically have a higher covering factor than their unobscured counterparts (Fig.\,\ref{fig:cfdust_vsNH}), consistently with previous studies of IR SEDs of nearby AGN (e.g., \citealp{Ramos-Almeida:2011eb}). This is in agreement with the idea that AGN with large covering factors of obscuring material are more likely to be observed as obscured/type-2, due to the larger fraction of obscured lines of sight \citep{Elitzur:2012dq}.
\item The dust covering factor shows a decrease with the Eddington ratio (\S\ref{sect:cfvslumeddratio}) similar to that observed in the X-ray band (\citealp{Ricci:2017ss,Ricci:2022xp}; Figs.,\ref{fig:CFdustvsEddratio_corr_taufive},\ref{fig:CFdust_IRvsXray}), which was attributed to the effect of radiation pressure on the dusty gas. The relation between the covering factor and $\lambda_{\rm Edd}$ implies that, besides the inclination angle, the observational classification of AGN into obscured and unobscured is driven by the Eddington ratio ({\it radiation-regulated unification model}; \citealp{Ricci:2017ss}). 
\item The obscuring material expelled by radiation pressure would be expected to populate the polar region of the system (see \S\ref{sec:polardust}), and could be responsible for the extended MIR emission observed in a growing number of AGN (e.g., \citealp{Honig:2012hq,Venanzi:2020dd}). In agreement with this, we find that all of the sources showing elongated mid-IR emission are accreting with Eddington ratios above the Eddington limit for dusty gas with $N_{\rm H}=10^{20}\rm\,cm^{-2}$, and more than half of them are above the Eddington limit for dusty gas with $N_{\rm H}=10^{22}\rm\,cm^{-2}$ (Fig.\,\ref{fig:cdf_polar}).
\item The median covering factors obtained for AGN located in different regions of the $N_{\rm H}-\lambda_{\rm Edd}$ plane is consistent with what would be expected if the growth of SMBHs were regulated by radiation (\citealp{Ricci:2022xp}; \S\ref{sect:RRgrowth}), and if AGN moved through the $N_{\rm H}-\lambda_{\rm Edd}$ plane during their lifetime (Fig.\,\ref{fig:RRgrowth_CF}).
\end{itemize}

\begin{acknowledgments}

We thank the referee for their comments, which helped us improve the quality of our manuscript. We acknowledge support from the National Science Foundation of China 11721303, 11991052, 12011540375 and 12233001 (LH) the National Key R\&D Program of China 2022YFF0503401 (LH), the China Manned Space Project CMS-CSST-2021-A04 and CMS-CSST-2021-A06 (LH), Fondecyt Regular grant 1230345 (CR) and ANID BASAL project FB210003 (CR, FEB, ET); NASA through ADAP award NNH16CT03C (MK); the Israel Science Foundation through grant number 1849/19 (BT); the European Research Council (ERC) under the European Union's Horizon 2020 research and innovation program, through grant agreement number 950533 (BT); Fondecyt fellowship No. 3220516 (MT); the Korea Astronomy and Space Science Institute under the R\&D program(Project No. 2022-1-868-04) supervised by the Ministry of Science and ICT (KO);  the National Research Foundation of Korea (NRF-2020R1C1C1005462) (KO); Fondecyt Postdoctorado 3210157 (AR); RIN MIUR 2017 project ''Black Hole winds and the Baryon Life Cycle of Galaxies: the stone-guest at the galaxy evolution supper", contract 2017PH3WAT (FR); the ANID - Millennium Science Initiative Program - ICN12\_009 (FEB), CATA-Basal - ACE210002 (FEB, ET), FONDECYT Regular - 1190818 (FEB, ET) and 1200495 (FEB, ET), N\'ucleo Milenio NCN\_058 (ET), the Science Fund of the Republic of Serbia, PROMIS 6060916, BOWIE and by the Ministry of Education, Science and Technological Development of the Republic of Serbia through the contract No.~451-03-9/2023-14/200002 (MS). This work made use of data from the NASA/ IPAC Infrared Science Archive and NASA/IPAC Extragalactic Database (NED), which are operated by the Jet Propulsion Laboratory, California Institute of Technology, under contract with the National Aeronautics and Space Administration. This research has made use of data and/or software provided by the High Energy Astrophysics Science Archive Research Center (HEASARC), which is a service of the Astrophysics Science Division at NASA/GSFC and the High Energy Astrophysics Division of the Smithsonian Astrophysical Observatory.

\end{acknowledgments}

\facilities Swift, WISE, IRAS, Akari, Spitzer

\bibliography{abs_bass2}
\bibliographystyle{aasjournal}

 \end{document}